\documentclass[aps,prd,showpacs,showkeys,superscriptaddress,twocolumn]{revtex4}
\usepackage{epsfig}
\usepackage{graphicx}
\usepackage{amsmath,amssymb,amsfonts,latexsym}
\usepackage{graphicx}

\usepackage{epsfig,amssymb,amsfonts,verbatim}

\usepackage{amsmath}
\usepackage{latexsym}
\usepackage{amsfonts}
\usepackage{amssymb}
\usepackage{color}

\def\bfl{\begin{flushleft}}
\def\efl{\end{flushleft}}
\def\bfr{\begin{flushright}}
\def\efr{\end{flushright}}
\def\bc{\begin{center}}
\def\ec{\end{center}}

\def\ba{\begin{eqnarray}}
\def\ea{\end{eqnarray}}
\def\baa#1{\begin{array}{#1}}
\def\eaa{\end{array}}
\def\bw{\begin{widetext}}
\def\ew{\end{widetext}}
\def\nn{\nonumber }

\def\text#1{\mbox{#1}}

\begin{document}


\title{Controlling electronic and adiabatic isolation of quantum dots from the substrate: An ionization-energy theoretic study}

\author{Andrew Das Arulsamy}
\email{andrew@physics.usyd.edu.au}
\affiliation{School of Physics, The
University of Sydney, Sydney, New South Wales 2006, Australia}

\author{Kostya (Ken) Ostrikov}
\affiliation{CSIRO Materials Science and Engineering, P.O. Box 218, Lindfield NSW 2070, Australia}
\affiliation{School of Physics, The University of Sydney, Sydney, New South Wales 2006, Australia}

\date{\today}

\begin{abstract}
Recent controversy on the quantum dots dephasing mechanisms (between pure and inelastic) is re-examined by isolating the quantum dots from their substrate by using the appropriate limits of the ionization energy theory and the quantum adiabatic theorem. When the phonons in the quantum dots are isolated adiabatically from the phonons in the substrate, the elastic or pure dephasing becomes the dominant mechanism. On the other hand, for the case where the phonons from the substrate are non-adiabatically coupled to the quantum dots, the inelastic dephasing process takes over. This switch-over is due to different elemental composition in quantum dots as compared to its substrate. We also provide unambiguous analyses as to understand why GaAs/AlGaAs quantum dots may only have pure dephasing while InAs/GaAs quantum dots give rise to the inelastic dephasing as the dominant mechanism. Our study accentuates the importance of the elemental composition (of both quantum dots and substrate) in evaluating the dephasing mechanisms of quantum dots. 
\end{abstract}


\maketitle

\section{Introduction}

The electron-phonon ($e$-$ph$) interaction in quantum dots (QDs) is evaluated theoretically by invoking the discontinuous dielectric property (between QDs and substrate) and quantum adiabatic approximation in order to understand the coupling between electrons in QDs and the phonons in both QDs and substrate. In order to do so, we propose here that the elemental composition of the QDs and the substrate need to be explicitly considered. We further show that such consideration will lead us to understand the dephasing mechanisms in QDs accurately for any non-free-electron QDs and substrate. Apparently, (1) the $e$-$ph$ interaction within the QDs and (2) between QDs and the substrate are crucial to the understanding of the dephasing mechanisms~\cite{bor}. Pure dephasing process is defined as due to the elastic interaction between electrons and phonons that only further corrects the unperturbed energy levels in QDs. Such a process does not excite the electrons or change the excited carrier density or the carrier occupation numbers~\cite{sang,skin}. Whereas, inelastic dephasing requires phonons to assist the electrons and/or holes to be relaxed to a lower energy-level, or electron relaxation due to phonon emission~\cite{mul,zib,sood}. Both pure- and inelastic-dephasing processes hinder the storage of quantum information, which are characterized by the timescales, $T_1$ and $T_2$, respectively~\cite{thomas}. The respective $T_1$ and $T_2$ timescales are also known as the phase and population relaxation lifetimes. 

Presently, there are several theories on these dephasing processes and are given in the Refs.~\cite{fom,poka,krum,gou,mul} that employ the adiabatic approximation. In those treatments however, the adiabaticity is invoked between the nuclear motion and the electronic excitation with respect to degeneracies, without incorporating the phonons from the substrate~\cite{fom,poka,krum,gou,mul}. Apart from that, a polaronic model based on the polaron relaxation was proposed to understand the relaxation mechanisms in QDs~\cite{grange}, which employs the approach of Klemens, Barman and Srivastava~\cite{kle,sri,sri2}. In their study, only Fr$\ddot{\rm o}$hlich polarons or the electron-LO(longitudinal optical) phonon interaction was considered. This model is an improved version of other reported relaxation models, namely, (1) the semi-classical approach~\cite{li} and (2) a phenomenological model that invoked the anharmonic mechanism for the bulk LO phonons within the Fermi golden rule framework~\cite{ver,jacc,vallee}. In all these approaches, (1) the origin of phonons is from the QDs, where the LO phonon contributions from the substrate or wetting layer are ignored by assuming the substrate phonons are the low-energy acoustic phonons, and (2) the phonons are independent of elemental composition.

Here, we take another step forward to understand the influence of substrate phonons (both acoustic and optical) on QDs dephasing mechanisms (both pure and inelastic) in detail. The influence of substrate phonons will be captured by studying the elemental compositions in QD and substrate materials separately. We will invoke the ionization energy theory (IET) to analyze how different elemental compositions in QDs and substrate will give rise to pure or inelastic dephasing processes or both. The reasons for incorporating the elemental compositions in QDs and substrate are (1) to investigate the reports that claim the matrix and/or substrate are indeed influencing both the electronic and phononic properties of the QDs~\cite{seong,nam,sun,masu,mogi}, and (2) to develop a strategy based on the IET to distinguish the contradicting results obtained by Sanguinetti et al.~\cite{sang} and others: Zibik et al.~\cite{zib,zibnat} and Chernikov et al.~\cite{chern}. For example, pure dephasing was found to be responsible in GaAs QDs on Al$_{0.3}$Ga$_{0.7}$As substrate~\cite{sang}. Whereas, inelastic dephasing played the major role in InAs QDs embedded in the GaAs matrix~\cite{zib,zibnat,chern}. Therefore, our primary intention here is to understand why and how different constituent atoms in QDs and substrate materials may contribute to the different electronic relaxation or dephasing mechanisms in QDs.

In order to achieve this, we will employ the IET, which has been developed earlier~\cite{arul1,arul2,arul3,arul4}. The reason to use the IET is because the theory is straightforward and the physical mechanisms derived from the IET can be directly related to the atomic constituents of any non-free-electron systems~\cite{ako1,ako2,ako3}. We organize the paper in the following order. A brief introduction to the IET is given in the following section. Subsequently, technical discussion are developed on the technique of isolating the QDs from its substrate. The QDs are first electronically isolated and followed by adiabatic isolation. Detailed discussion on experimental proofs and the possible applications of the IET are highlighted with predictions. In addition, we also explain how the different dephasing rates due to defects, impurities, electron-electron and electron-phonon interactions, and spins are related to the ionization energy concept. 

\section{Elements of the ionization energy theory}

We start from the many-body Hamiltonian, which is given by~\cite{arul1,arul2} 

\begin {eqnarray}
\hat{H}\varphi = (E_0 \pm \xi)\varphi, \label{eq:2}
\end {eqnarray}

where, the eigenvalue is exactly equals to $E_0 \pm \xi$. Here, $E_0$ is the total energy of the system at zero temperature ($T = 0$ K), $\xi$ is the energy-level difference in a given atom or the ionization energy. Importantly, the IET presented here can be used to study any non-free-electron materials. If a given material is a free-electron metal, then Eq.~(\ref{eq:2}) reduces to the standard time-independent Schr$\ddot{\rm o}$dinger equation given by, $\hat{H}\varphi = E\varphi$, where one needs other theoretical and computational methods to solve it by means of variational principle~\cite{mxu}. For solids however, the parameter $\xi$ in the total energy, $E_0 \pm \xi$ refers to the energy level differences in solids, which is difficult to be determined. Therefore, we will use the ionization energy approximation, $\xi \propto E_I$, where $E_I$ is the atomic ionization energy (for an isolated or free atom). Mathematically, the ionization energy approximation can be written as 

\begin {eqnarray}
E_0 \pm \xi \propto E_0 \pm \sum_i^z \frac{E_{Ii}}{z}, \label{eq:6}
\end {eqnarray}

where, the subscript $i$ counts the first, second, ..., $z$ ionization energy of each constituent atom for a given material. Here, $\sum_i^z E_{Ii}/z$ gives the changes to the average ionization energy of a given system. In addition, in this approximation, one needs to rely on the accuracy of the ionic valence states of the constituent atoms and as such, knowing accurate valence state values are important, and usually, they are easily predictable. For example, these valence states can be obtained from their stable oxidation states and these states are also known to vary significantly as a result of defect-formation and/or other structural deformations~\cite{arul3}. Subsequently, we can substitute the new total energy, $E_0 \pm E_I$ (after the approximation) into the ionization energy based Fermi-Dirac statistics~\cite{arul1,arul2} as given below

\begin{eqnarray}
&&f_e(E_0,E_I) = \frac{1}{e^{[\left(E_0 + E_I
\right) - E_F^{(0)}]/k_BT }+1}, \nn \\&& f_h(E_0,E_I) = \frac{1}{e^{[E_F^{(0)} - \left(E_0 - E_I
\right)]/k_BT}+1}. \label{eq:14}
\end{eqnarray} 

Note here that for QDs, $E_F^{(0)}$ can be regarded as the highest occupied energy-level at $T$ = 0 K, and $k_B$ is the Boltzmann constant. Furthermore, $f_e(E_0,E_I)$ and $f_h(E_0,E_I)$ are the probability functions for electrons and holes, respectively. Even though Eq.~(\ref{eq:2}) is for $T = 0$ K, the temperature effect can be taken into account via Eq.~(\ref{eq:14}). Note here that the appearance of $\xi$ in any equations simply means that the ionization energy approximation has not been invoked. After applying the approximation, one must replace the $\xi$ with $E_I$ in all the equations. This approximation will be used later to calculate the average ionization energy of the elements present in a given QD. The size effect is easily obtained from a simple one-dimensional infinite square-well potential, which is given by~\cite{ako3} 

\begin {eqnarray}
a = \frac{\hbar n \pi}{\sqrt{2m(E_0 \pm \xi_a)_n}}. \label{eq:13} 
\end {eqnarray}

Here, $a$ is the width of the potential well and we can see the inverse proportionality between $a$ and $\xi$, while $m$, $\hbar$ and $n$ denote the electronic mass, Planck constant and the principal quantum number, respectively. Importantly, $E_I$ is unique for each atom that will assist us to capture the effect of different constituent atoms in QD and substrates on their electronic excitation probability and the electron-phonon interaction. Therefore, the only input parameter is the type of atoms or ions that may exist in a given sample, and their appropriate valence states. More details about the IET and its approximation can be found in Refs.~\cite{arul1,arul3,arul2,arul4}. In the subsequent sections however, we will use $\xi$ throughout instead of $E_I$ for notational clarity. Thus, any values calculated for $\xi$ implies that we have invoked the ionization energy approximation. 

\section{Electronic and adiabatic isolation of QDs from the substrate}

In the following sections, the electronic interaction between the arrays of QDs and the substrate is evaluated with respect to the possibility of isolating the electronic property of QDs from the substrate. After that, these QDs are separately (1) coupled to QD and substrate phonons, and (2) isolated from those phonons adiabatically. This adiabatic approach is used to understand the influence of the substrate and QD phonons with QD electrons. We also compare our theoretical results with recent experimental observations.  

\subsection{Electronically isolated QDs from the substrate}

Let us first evaluate the charge distribution of the QDs with respect to its substrate using the Gauss and Green's theorems. From the Gauss theorem~\cite{smy}, we can write 

\begin{eqnarray}
\sum_{j=1}^m\int_{S_j}\vec{A} \cdot n_j ~dS_j = \int_{\Omega}\nabla \cdot \vec{A} ~d\Omega, \label{eq:15}
\end{eqnarray} 

\begin{figure}[hbtp!]
\begin{center}
\scalebox{0.25}{\includegraphics{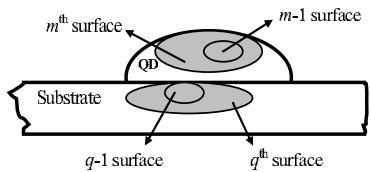}}
\caption{Two-dimensional schematic representation of an isolated QD and substrate surfaces with $m-1$ and $q-1$ closed surfaces, enclosed by the $m^{\rm{th}}$ and $q^{\rm{th}}$ surfaces, respectively.}
\label{fig:1}
\end{center}
\end{figure}

where, $\Omega$, $n_j$ and $\vec{A}$ denote the volume, unit vector normal to the surface and any given vector, respectively with $j$ = 1, 2,... $m$ counts the number of closed surfaces (see Fig.~\ref{fig:1}). We can now use Eq.~(\ref{eq:15}) by first defining, $\vec{A}$ = $\vec{D}$ = $\epsilon\vec{E}$ = $-\epsilon\nabla V$, $V$, $\vec{D}$ and $\vec{E}$ denote the electrostatic potential, charge displacement and the electric field, respectively. Subsequently, by substituting them into Eq.~(\ref{eq:15}) we can obtain

\begin{eqnarray}
\sum_{j=1}^m\int_{S_j}\epsilon\nabla V \cdot n_j dS_j = \int_{\Omega}\nabla \cdot \epsilon\nabla V d\Omega = \int_{\Omega} \epsilon\nabla^2 Vd\Omega. \label{eq:16}
\end{eqnarray}

Thus far, we have not explicitly described the quantity, $\epsilon$, which is a dielectric constant. However, $\epsilon$ is obviously discontinuous at the boundary between the substrate and QDs. For a given array of QDs on a particular substrate, there will be $m-1$ surfaces enclosed by the $m^{\rm{th}}$ surface and similarly, there will be $q-1$ surfaces enclosed by the $q^{\rm{th}}$ surface for the substrate (SUB). The purpose of writing Eq.~(\ref{eq:16}) is to obtain separate charge distributions in QDs and substrate. By isolating them, one can further evaluate the strength of charge accumulation in QDs, as compared to their substrate. As such, one can rewrite Eq.~(\ref{eq:16}) to arrive at 

\begin{eqnarray}
&&\sum_{j=1}^m\int_{S_j}\epsilon^{\rm{SUB}}\bigg(\frac{\partial V^{\rm{SUB}}_j}{\partial n_j}\bigg) dS_j - \sum_{p=1}^q\int_{S_p}\epsilon^{\rm{QD}}\bigg(\frac{\partial V^{\rm{QD}}_p}{\partial n_p}\bigg) dS_p \nn \\&& = \int_{\Omega}[\nabla \cdot \vec{A}^{\rm{SUB}} - \nabla \cdot \vec{A}^{\rm{QD}}] d\Omega \nn \\&& = \int_{\Omega} [\nabla \cdot (\epsilon^{\rm{SUB}}\nabla V^{\rm{SUB}}) - \nabla \cdot (\epsilon^{\rm{QD}}\nabla V^{\rm{QD}})]d\Omega \nn \\&& = \int_{\Omega}[\epsilon^{\rm{SUB}} \nabla^2 V^{\rm{SUB}} - \epsilon^{\rm{QD}} \nabla^2 V^{\rm{QD}}]d\Omega, \label{eq:17}
\end{eqnarray}
  
where, $dS_j$ is an element of the substrate surface, and $dS_p$ is for QD surface. For QD arrays on a substrate, we will consider three conditions, namely,

\begin{eqnarray}
&&\xi^{\rm{QD}} \ll \xi^{\rm{SUB}}~~~~~~~~\rm{condition~(i)}, \nn \\&&
\xi^{\rm{SUB}} \ll \xi^{\rm{QD}}~~~~~~~~\rm{condition~(ii)}, \nn \\&&
\xi^{\rm{QD}} \approx \xi^{\rm{SUB}}~~~~~~~~\rm{condition~(iii)}. \nn
\end{eqnarray}

These conditions will be used to identify whether the electrons in QDs are both electronically and/or adiabatically isolated from their substrate. Satisfying condition (i) implies that one can use the last integral in Eq.~(\ref{eq:17}) to obtain (after using Poisson equation, $\epsilon\nabla^2V = -\rho/\epsilon_0$)

\begin{eqnarray}
\left|\int_{\Omega} \frac{\rho(\textbf{r})^{\rm{SUB}}_{\rm{induced}}}{\epsilon_0}~d\Omega\right| \ll \left|\int_{\Omega} \frac{\rho(\textbf{r})^{\rm{QD}}_{\rm{induced}}}{\epsilon_0}~d\Omega\right|, \label{eq:18}
\end{eqnarray} 

where, $\rho(\textbf{r})$ is the charge density, which is given by~\cite{arul2}

\begin{eqnarray}
&\rho(\textbf{r})& = \rho_0(\textbf{r}) + \rho(\textbf{r})_{\rm{ind}} \nn \\&& = n_0e + \frac{3n_0e^2}{2E_F^0}V(\textbf{r})e^{\lambda(E_F^0 - \xi)}, \label{eq:19}
\end{eqnarray}

where $e$ is the electronic charge, while $\lambda = (12\pi\epsilon_0/e^2)a_B$, in which $a_B$ is the Bohr radius and $\epsilon_0$ denotes the permittivity of free space. Note here that $\rho_0(\textbf{r})$ and $n_0$ are the charge and carrier densities, respectively at $T = 0$ K and without any external disturbances. Whereas, the induced charge density, $\rho(\textbf{r})_{\rm{ind}}$ capture the charge accumulation due to temperature, $T > 0$ K and due to other external disturbances. The potential, $V(\textbf{r}) = V(\textbf{r})_{\rm{ext}} + V(\textbf{r})_{\rm{ind}}$, where $V(\textbf{r})_{\rm{ext}}$ is the external potential. If $\rho(\textbf{r})_{\rm{ind}}$ = 0, then we will arrive at the Green's reciprocation theorem~\cite{smy}. For example, after letting $p$ = $j$ and $q$ = $m$, so that, $\Omega^{\rm{SUB}} = \Omega^{\rm{QD}}$ (for mathematical convenience)

\begin{eqnarray}
&&\sum_{j=1}^m\int_{S_j}\epsilon^{\rm{SUB}}\bigg(\frac{\partial V^{\rm{SUB}}}{\partial n_j}\bigg) dS_j \nn \\&& -\sum_{j=1}^m\int_{S_j}\epsilon^{\rm{QD}}\bigg(\frac{\partial V^{\rm{QD}}}{\partial n_j}\bigg) dS_j  \approx 0, \label{eq:20}
\end{eqnarray}    

or equivalently

\begin{eqnarray}
\sum_{j=1}^m\bigg[\int_{S_j}\gamma^{\rm{SUB}}_j~dS_j ~- \int_{S_j}\gamma^{\rm{QD}}_j~dS_j\bigg] \approx 0, \label{eq:21}
\end{eqnarray}    

where $\gamma$ denotes the surface charge density and subsequently, we can arrive at

\begin{eqnarray}
\sum_{j=1}^m e_j^{\rm{QD}} \approx \sum_{j=1}^m e_j^{\rm{SUB}}, \label{eq:22}
\end{eqnarray}    
   
because, $\gamma^{\rm{SUB}} \approx \gamma^{\rm{QD}}$ or $\rho_0^{\rm{SUB}}(\textbf{r}) \approx \rho_0^{\rm{QD}}(\textbf{r})$, which is to say, both QDs and the substrate are approximately uncharged in the absence of temperature and external disturbances. Now, we can repeat this procedure exactly by only switching the indices, QD and SUB to obtain condition (ii), $\xi^{\rm{SUB}} \ll \xi^{\rm{QD}}$. Hence, both conditions (i) and (ii) imply that the QDs are electronically isolated from the substrate. However, condition (iii), $\xi^{\rm{QD}} \approx \xi^{\rm{SUB}}$ gives rise to $e$-$e$ interaction between QDs and its substrate. For example, from condition (i), we have $\xi^{\rm{QD}} \ll \xi^{\rm{SUB}}$ $\rightarrow$ $\omega_e^{\rm{QD}} \ll \omega_e^{\rm{SUB}}$ $\rightarrow$ $f_e(E_0,E_I)^{\rm{QD}} \gg f_e(E_0,E_I)^{\rm{SUB}}$, thus, the characteristic electronic (subscript $e$) excitation probability for QDs is much higher compared to its substrate, and vice versa for condition (ii) (for example, see Eq.~(\ref{eq:14})). Note here that $\xi^{\rm{QD}}/\hbar = \omega_e^{\rm{QD}}$ and $\xi^{\rm{SUB}}/\hbar = \omega_e^{\rm{SUB}}$. 

Where condition (iii) holds, the electronic properties of the substrate will also affect QDs and vice versa. This means that the electronic excitation probability for both QDs and the substrate are identical and therefore they (QDs and the substrate) are electronically coupled. In other words, condition (iii) also implies that the confinement energy is close to zero. However, this electronic coupling does not occur in real QDs that have finite confinement energies. This may explain why the QDs are always electronically isolated if the confinement energy $\neq$ 0. Shapes of QD structures also significantly affect their electronic properties~\cite{kaya,le,jit,santos}. This shape dependence will give rise to anisotropic ionization energy, and becomes important, if and only if we are comparing QDs of different shapes. In reality, for a given nanoscale synthesis process, shapes of QDs are quite similar, whereas size non-uniformities are quite substantial~\cite{kostya1,kostya2}. In cases where QDs have different shapes, $\xi$ can be spatially averaged by incorporating the QD size, $a$ correctly (will be a complicated algebraic function) in Eq.~(\ref{eq:13}).    

\subsection{Electron-phonon interaction and adiabatic-isolation of QDs from the substrate}

\begin{figure}[hbtp!]
\begin{center}
\scalebox{0.3}{\includegraphics{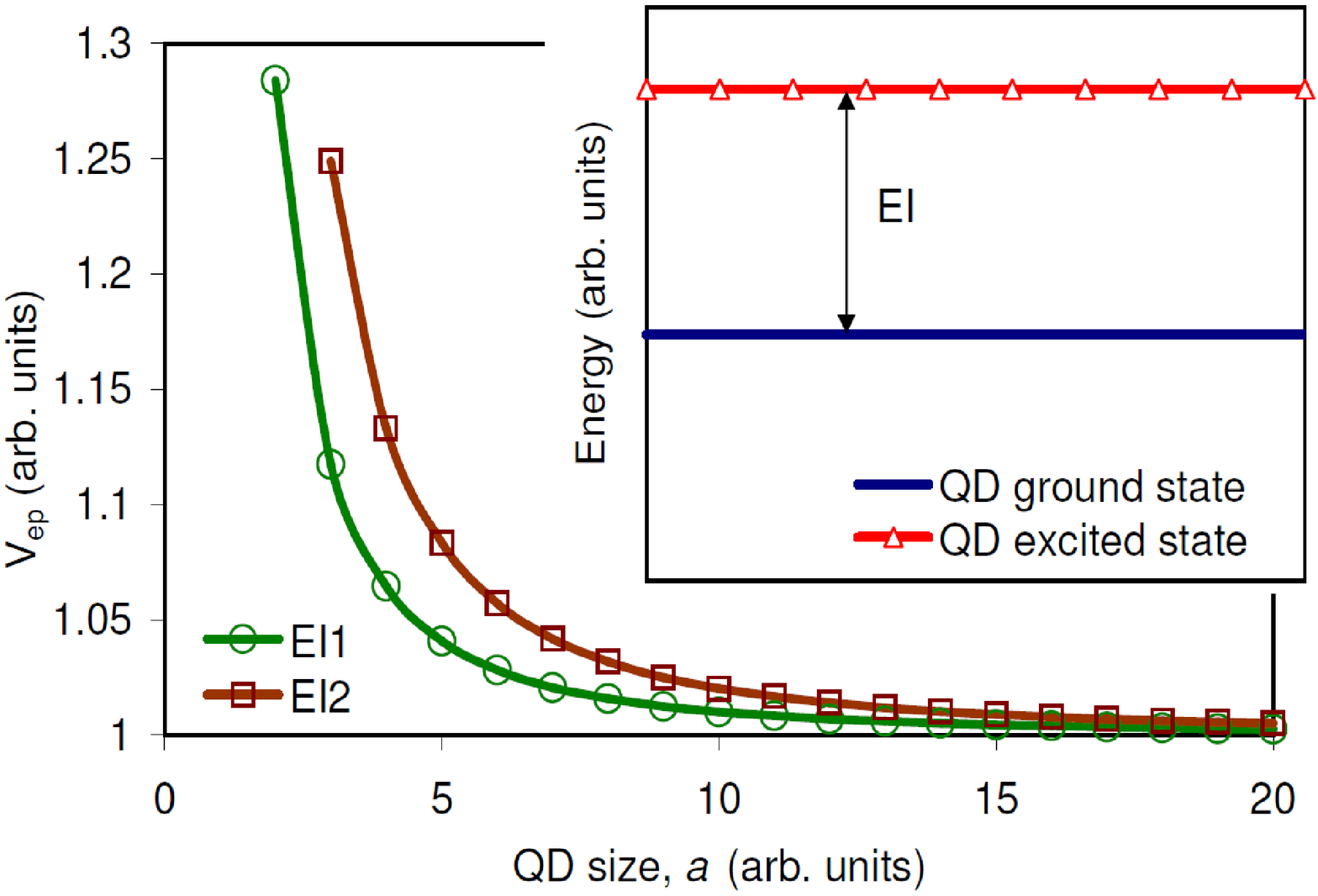}}
\caption{Electron-phonon interaction energy as a function of QD size, $a$ for two different ionization energy values, $E_I1 < E_I2$. These plots were obtained from Eqs.~(\ref{eq:25}) and~(\ref{eq:13}). The INSET shows the definition of the ionization energy in a QD, with appropriate labels for the ground and excited state energies. Note here that the ionization energy is the energy-level difference between these two states.}
\label{fig:2}
\end{center}
\end{figure}

The next issue here is to understand the phonon contribution to the electronic properties of QDs. We will investigate the possibility of isolating the substrate phonons from QDs, or coupling them to QDs. The motivation to study these electronic (given earlier) and adiabatic isolations are important so as to evaluate the dephasing mechanisms namely, elastic (pure) and inelastic dephasing. We will address both of these mechanisms separately here with the IET and the quantum adiabatic approximation. It has been shown in our earlier work, that the phonon frequencies of both optical and acoustic, for 1D diatomic system with two different ions (masses, $M_1$ and $M_2$) is given by~\cite{arul1}

\begin {eqnarray}
&& w^2_{\pm} = \frac{e^{\lambda(\xi - E_F^0)}}{2M_1M_2} \big[A \pm B^{\frac{1}{2}}\big], \label{eq:24} \\&& A = (Q + G)(M_1 + M_2), \nn \\&&  
B = (Q + G)^2 (M_1 + M_2)^2 - \nn \\&& 4M_1M_2[(Q + G)^2 - Q^2 - G^2 - 2QG\cos(\textbf{k}\bar{u})].\nn
\end {eqnarray}

where, $Q$ and $G$ are the interaction potential constants, $\bar{u}$ is a lattice site, whereas, the screened $e$-$ph$ Coulomb potential is given by~\cite{arul1}

\begin {eqnarray}
V_{\rm{ep}}(\textbf{k},\textbf{k}^*) = \frac{1}{\Omega\epsilon_0}\bigg[\frac{e^2}{|\textbf{k}-\textbf{k}^*|^2 + K_s^2\exp[\lambda(E_F^0 - \xi)]}\bigg],\label{eq:25}
\end {eqnarray} 

which is in exact form, with the screened $e$-$e$ Coulomb potential~\cite{arul2},

\begin {eqnarray}
V_{\rm{ee}}(\textbf{k}) = \frac{1}{\Omega\epsilon_0}\bigg[\frac{e^2}{\textbf{k}^2 + \textsl{K}_s^2\exp[\lambda(E_F^0 - \xi)]}\bigg], \label{eq:26}
\end {eqnarray}

where, $\textbf{k} - \textbf{k}^* = \textbf{q}$ is due to the crystal momentum conservation. Here, $\textbf{k}$ and $\textit{K}_s$ denote the wavevector and Thomas-Fermi wavenumber, respectively. Figure~\ref{fig:2} shows the $V_{\rm{ep}}$ dependence on the ionization energy and the QD size, $a$ that follows from Eqs.~(\ref{eq:25}) and~(\ref{eq:13}), while the inset captures the definition of the ionization energy in a given QD. Note here that the ionization energy here implies the energy-level difference between the ground and excited states. 

From Eq.~(\ref{eq:24}), condition (i) implies $w_{\rm{ph}}^{\rm{QD}} \ll w_{\rm{ph}}^{\rm{SUB}}$ $\rightarrow$ $t^{\rm{QD}} \gg t^{\rm{SUB}}$, where, $t^{\rm{QD}}$ and $t^{\rm{SUB}}$ denote the characteristic timescales (\textit{not the relaxation lifetimes or dephasing rates}) for the phonons in QDs and substrate, respectively. Note here that $w_{\rm{ph}}$ can be due to optical or acoustic or both.  

\subsubsection{Analysis for $w_{\rm{ph}}^{\rm{QD}} \ll w_{\rm{ph}}^{\rm{SUB}}$}

\textbf{Claim 1}: The electrons in QDs can be made adiabatically coupled to the substrate or QD phonons or both by employing condition (i). 

\textbf{Proof 1}: From condition (i), we have $\xi^{\rm{QD}} \ll \xi^{\rm{SUB}}$ that implies $w_{\rm{ph}}^{\rm{QD}} \ll w_{\rm{ph}}^{\rm{SUB}}$ because $\xi^{\rm{QD,SUB}} \propto w_{\rm{ph}}^{\rm{QD,SUB}}$ (from Eq.~(\ref{eq:24})). The next step is to identify the relationship between $\xi^{\rm{QD}}$ and $w_{\rm{ph}}^{\rm{QD,SUB}}$, where we can define four inequalities, \\

A1~~ $\xi^{\rm{QD}} > \hbar w_{\rm{ph}}^{\rm{SUB}}$ $\Rightarrow$ $e^{\rm{QD}}$-$ph^{\rm{SUB}}$ $\Rightarrow$ $t^{\rm SUB} > \tau^{\rm QD}$, \\

B2~~ $\xi^{\rm{QD}} > \hbar w_{\rm{ph}}^{\rm{QD}}$ $\Rightarrow$ $e^{\rm{QD}}$-$ph^{\rm{QD}}$ $\Rightarrow$ $t^{\rm QD} > \tau^{\rm QD}$, \\

C3~~ $\xi^{\rm{QD}} < \hbar w_{\rm{ph}}^{\rm{SUB}}$ $\Rightarrow$ $e^{\rm{QD}}$-$ph^{\rm{SUB}}$ $\Rightarrow$ $t^{\rm SUB} < \tau^{\rm QD}$, \\ 

D4~~ $\xi^{\rm{QD}} < \hbar w_{\rm{ph}}^{\rm{QD}}$ $\Rightarrow$ $e^{\rm{QD}}$-$ph^{\rm{QD}}$ $\Rightarrow$ $t^{\rm QD} < \tau^{\rm QD}$, \\

where $\tau^{\rm QD}$ is the electronic relaxation lifetime or dephasing rate in the QDs. Both A1 and B2 will give rise to pure dephasing because the phonons timescales from the substrate ($t^{\rm SUB}$) and QDs ($t^{\rm QD}$) are low energy phonons compared to the QD energy-level difference ($\xi^{\rm{QD}}$). Invoking condition (i) implies A1 and B2 are unlikely and thus pure dephasing is negligible. On the other hand, the inequalities in C3 and D4 satisfy condition (i) and will lead us to inelastic dephasing because of the adiabatic coupling between the electrons in the QDs and phonons from both QDs and substrate. Hence, there is always a finite probability for the absorption and emission of QD ($e^{\rm{QD}}$-$ph^{\rm{QD}}$) and/or substrate ($e^{\rm{QD}}$-$ph^{\rm{SUB}}$) phonons $\blacksquare$ 

Now, we know that condition (i) gives rise to inelastic dephasing as the dominant mechanism. Therefore, the relaxation lifetime, $\tau^{\rm QD}$ can only be made longer by further enforcing the inequalities $t^{\rm QD} < \tau^{\rm QD}$ and/or $t^{\rm SUB} < \tau^{\rm QD}$. In other words, we need to satisfy and enforce the inequalities, $\xi^{\rm{QD}} < \hbar w_{\rm{ph}}^{\rm{QD}}$ and/or $\xi^{\rm{QD}} < \hbar w_{\rm{ph}}^{\rm{SUB}}$. For example, enforcing the inequality, $\xi^{\rm{QD}} < \hbar w_{\rm{ph}}^{\rm{QD}}$ means $\hbar w_{\rm{ph}}^{\rm{QD}} - \xi^{\rm{QD}} \rightarrow \infty$. This is exactly what have been done by Zibik et al.~\cite{zibnat}. 

For example, the inequality, C3 (and/or D4) have been achieved recently by deliberately changing the elemental composition of InGaAs/GaAs QDs via annealing~\cite{zibnat}. Note here that $\xi_{\rm InGaAs}^{\rm{QD}} < \xi_{\rm GaAs}^{\rm{SUB}}$. They have (a) obtained the inelastic dephasing due to C3 and/or D4 satisfying the theoretical results discussed above, and (b) achieved longer relaxation lifetime in QDs ($\tau^{\rm QD}$) by enforcing $\xi^{\rm{QD}} < \hbar w_{\rm{ph}}^{\rm{SUB,QD}}$. In other words, they systematically reduce the photoluminescence (PL) energy peaks by annealing their samples. Reducing PL energy peaks means decreasing $\xi^{\rm{QD}}$ and therefore increasing the magnitude of $\hbar w_{\rm{ph}}^{\rm{QD}} - \xi^{\rm{QD}}$. As a result, they have enforced the inequalities, $t^{\rm QD} < \tau^{\rm QD}$ and/or $t^{\rm SUB} < \tau^{\rm QD}$ as predicted by the IET and condition (i).     

\subsubsection{Analysis for $\xi^{\rm{QD}} \gg \xi^{\rm{SUB}}$}

\textbf{Claim 2}: The electrons in QDs can also be made adiabatically independent from the substrate phonons by employing condition (ii), which is $\xi^{\rm{QD}} \gg \xi^{\rm{SUB}}$. In this case, we have $w_{\rm{ph}}^{\rm{QD}} \gg w_{\rm{ph}}^{\rm{SUB}}$ $\rightarrow$ $t^{\rm{QD}} \ll t^{\rm{SUB}}$ and therefore, the $e^{\rm{QD}}$-$ph^{\rm{SUB}}$ interaction between QD electrons and the substrate phonons can be assumed to be switched-off (no inelastic dephasing). To see this conclusion clearly, we need to recall proof 1 given earlier.  

\textbf{Proof 2}: After invoking condition (ii), C3 and D4 are unlikely and therefore the inelastic dephasing is negligible. Consequently, we are left with the inequalities given in A1 and B2, which satisfy condition (ii) and pure dephasing as the dominant mechanism $\blacksquare$

Here, the relaxation lifetime, $\tau^{\rm QD}$ can only be made longer by further enforcing the inequalities $t^{\rm SUB} > \tau^{\rm QD}$ and/or $t^{\rm QD} > \tau^{\rm QD}$ as given in A1 and B2. Based on condition (ii), we need to enforce the inequalities, $\xi^{\rm{QD}} > \hbar w_{\rm{ph}}^{\rm{QD}}$ and/or $\xi^{\rm{QD}} > \hbar w_{\rm{ph}}^{\rm{SUB}}$ with appropriate changes to the elemental composition of the QDs or substrate or both.

\subsubsection{Analysis for $w_{\rm{ph}}^{\rm{QD}} \approx w_{\rm{ph}}^{\rm{SUB}}$}

Finally, if the QDs were to satisfy condition (iii), then these QDs are just dots with no electronic confinement effect and however, their $e$-$ph$ interaction cannot be understood without additional complications. In this case, the ionic mass will start to play a crucial role as given in Eq.~(\ref{eq:24}). In other words, condition (iii) does not guarantee the relation, $w_{\rm{ph}}^{\rm{QD}} = w_{\rm{ph}}^{\rm{SUB}}$, and the reduced mass of the QDs and substrate will determine the correct relationship. For example, the reduced mass ($M_r$) is defined as $1/M_r = \sum_i 1/M_i$ and even if condition (iii) is fulfilled, we may still have $M_r^{\rm{QD}} \neq M_r^{\rm{SUB}}$, which will lead to $w_{\rm{ph}}^{\rm{QD}} \neq w_{\rm{ph}}^{\rm{SUB}}$. To understand this effect clearly, let us first assume that the QDs and the substrate are of same materials and elemental composition, and we do not intend to change the elemental composition. This assumption implies, $M_r^{\rm{QD}} = M_r^{\rm{SUB}}$ and the inequality, $w_{\rm{ph}}^{\rm{QD}} > w_{\rm{ph}}^{\rm{SUB}}$ is due to $\xi^{\rm{QD}} > \xi^{\rm{SUB}}$, which is entirely due to size. In addition, in this hypothetical case, condition (iii) can be achieved by making the size of the of QDs larger. By doing so, these QDs lose both their electronic and phononic confinements. 

On the other hand, if the QDs and the substrate materials are different ($M_r^{\rm{QD}} \neq M_r^{\rm{SUB}}$), then condition (iii) can be achieved by changing the elemental composition of the QDs. As a consequence, we have both $M_r^{\rm{QD}} \neq M_r^{\rm{SUB}}$ and $\xi^{\rm{QD}} = \xi^{\rm{SUB}}$, which imply that the QDs are not electronically isolated, but are still isolated with respect to phonons. In this second scenario, it is essential to note that one cannot vary $\xi^{\rm{QD}}$ independent of $M_r^{\rm{QD}}$. Apart from that, small changes in the reduced mass does not play a significant role in the presence of condition (i) and (ii) because the strength of the electronic polarizability for each ion is more effective in determining the $w_{\rm{ph}}$. This electronic polarizability is related to the interaction potential constants, $Qe^{\lambda(\xi - E_F^0)}$ and $Ge^{\lambda(\xi - E_F^0)}$, which takes the deformable-ion effect into account. Therefore, if we have a situation where $M_r^{\rm{QD}} > M_r^{\rm{SUB}}$ $\rightarrow$ $w_{\rm{ph}}^{\rm{QD}} < w_{\rm{ph}}^{\rm{SUB}}$ and $\xi^{\rm{QD}} > \xi^{\rm{SUB}}$ $\rightarrow$ $w_{\rm{ph}}^{\rm{QD}} > w_{\rm{ph}}^{\rm{SUB}}$, then $M_r$ and $\xi$ compete with each other. In this case, $\xi$ is more sensitive than $M_r$ because the ionization energy that control the rigidity (or deformability) of the ions determines the $w_{\rm{ph}}$. For example, slight changes in the reduced mass can change the phonon frequency, but not as effective as the ionization energy that control the ions deformability. However, any huge changes to $M_r$ will significantly reduce the effect of the ionization energy on $w_{\rm{ph}}$.   

\section{Experimental proofs and applications}

In this section we will scrutinize QDs made of two different materials in which, one of them has been studied by Sanguinetti et al.~\cite{sang} while the other one was investigated by Zibik et al.~\cite{zib}. Note that the QDs and their respective substrates discussed here have Zincblende crystal structure~\cite{wyc} and therefore, Eqs.~(\ref{eq:24}),~(\ref{eq:25}) and~(\ref{eq:26}) can be applied directly~\cite{arul1}. 

\subsection{GaAs QDs on Al$_{0.3}$Ga$_{0.7}$As substrate}

The former group~\cite{sang} studied GaAs QDs on Al$_{0.3}$Ga$_{0.7}$As substrate grown using modified-droplet epitaxy with QD density of the order of 10$^{8}$ cm$^{-2}$. They observed that the exciton dephasing of the QDs is independent of confinement energy with no quantum-size effect. This conclusion was based on their temperature-dependent PL spectra measurements of which, the confinement-energy ($\approx$ 160 meV) of one of their smaller QD sample, is much larger than the calculated activation energy ($\approx$ 30 meV, related to GaAs longitudinal optical (LO) phonon). Therefore, the LO phonons do not contribute to inelastic dephasing~\cite{sang}. As such, pure dephasing is the likely process with GaAs LO phonons. These low-energy LO phonons do not cause electron relaxation from an excited state to the ground state. 

Let us now apply our theory described earlier using conditions (i) and (ii) to GaAs QDs on Al$_{0.3}$Ga$_{0.7}$As substrate. The average ionization energies for both Ga and As are given by, $\xi_{\rm{Ga^{3+}}}$ = (578.8 + 1979.3 + 2963)/3 = 1840 kJmol$^{-1}$ and $\xi_{\rm{As^{3+}}}$ = (947 + 1798 + 2735)/3 = 1827 kJmol$^{-1}$, respectively. Prior to averaging, all the ionization energies mentioned in this work were obtained from Ref.~\cite{web}. This gives the total value of $\xi^{\rm{TOT}}_{\rm{QD}}$ = 1827 + 1840 = 3667 kJmol$^{-1}$. Similarly, the total average for the substrate can be calculated as, $\xi_{\rm{Ga^{3+}}}$ = [0.7$\times$(578.8 + 1979.3 + 2963)]/3 = 1288 kJmol$^{-1}$, $\xi_{\rm{Al^{3+}}}$ = [0.3$\times$(577.5 + 1816.7 + 2744.8)]/3 = 514 kJmol$^{-1}$, therefore, the total value for the substrate is $\xi^{\rm{TOT}}_{\rm{SUB}}$ = 1288 + 1827 + 514 = 3629 kJmol$^{-1}$. Thus, the difference is, $\xi^{\rm{TOT}}_{\rm{QD}}$ $-$ $\xi^{\rm{TOT}}_{\rm{SUB}}$ = 3667 $-$ 3629 = 38 kJmol$^{-1}$ = 400 meV atom$^{-1}$ and we have $\xi^{\rm{TOT}}_{\rm{SUB}}$ $<$ $\xi^{\rm{TOT}}_{\rm{QD}}$ $\rightarrow$ $w_{\rm{ph}}^{\rm{SUB}}$ $<$ $w_{\rm{ph}}^{\rm{QD}}$ $\rightarrow$ $t^{\rm{SUB}}$ $>$ $t^{\rm{QD}}$. Obviously, this satisfies condition (ii) and gives rise to adiabatically independent QDs and elastic dephasing is expected from our theory, in accordance with the experimental results and interpretations given in Ref.~\cite{sang}. 

Apart from that, the inequality, $\xi^{\rm{TOT}}_{\rm{SUB}}$ $<$ $\xi^{\rm{TOT}}_{\rm{QD}}$ implies that the electrons in the substrate has a higher probability to be excited at any given temperature, compared to the electrons in QDs. As a consequence, the substrate ions deformability is larger due to large electronic excitation probability, which in turn, explains why the lattice vibrational frequency for the QDs is larger than the substrate~\cite{arul1}. This means that, the ions in the QDs are more rigid compared to their substrate. 

Now, using the reduced mass (in atomic mass unit) as the main argument, we can obtain $M_r^{\rm{SUB}}$ (28.98) $<$ $M_r^{\rm{QD}}$ (36.11) $\rightarrow$ $w_{\rm{ph}}^{\rm{SUB}}$ $>$ $w_{\rm{ph}}^{\rm{QD}}$, which is in contradiction with the analysis stated above. However, as explained earlier, we have assumed that the effect of the small reduced mass is not as substantial as the ionization energy effect in accordance with Ref.~\cite{arul1}. Physically, what this means is that when the change in the reduced mass competes with the change in the ionization energy, then for as long as the change in the reduced mass is relatively small, the ionization energy effect will stay significant.      

\subsection{InAs QDs in GaAs matrix}

The second evidence comes from the work of Zibik et al.~\cite{zib}. Their QDs are made from InAs grown layer-by-layer with GaAs as the matrix and barrier. They used the four-wave-mixing (FWM) spectroscopy to study the dephasing mechanism in QDs. The samples were grown using the molecular beam epitaxy in the Stranski-Krashtanow mode and the QDs were separated by 50 nm wide GaAs barriers to prevent structural and electronic coupling between the QD layers~\cite{zib}. Both their calculations and experimental results point toward intersublevel (inelastic) dephasing in electron-doped QDs, which revealed oscillatory behavior for the polarization decay (for times $<$ 5 ps). By repeating our $\xi$ averaging as shown above, and after identifying the GaAs matrix/barrier as the substrate, we can arrive at $\xi^{\rm{TOT}}_{\rm{SUB}}$ $-$ $\xi^{\rm{TOT}}_{\rm{QD}}$ = 3667 $-$ 3521 = 146 kJmol$^{-1}$ = 1500 meV atom$^{-1}$. In this case, we have $\xi^{\rm{TOT}}_{\rm{SUB}}$ $>$ $\xi^{\rm{TOT}}_{\rm{QD}}$ $\rightarrow$ $w_{\rm{ph}}^{\rm{SUB}}$ $>$ $w_{\rm{ph}}^{\rm{QD}}$ $\rightarrow$ $t^{\rm{SUB}}$ $<$ $t^{\rm{QD}}$. Hence, InAs QDs satisfy condition (i) in which, the phonons from GaAs coupled to QDs non-adiabatically and favor the inelastic dephasing mechanism. 

The inequality, $\xi^{\rm{TOT}}_{\rm{SUB}}$ $>$ $\xi^{\rm{TOT}}_{\rm{QD}}$ also implies that the electrons in the substrate have a lower probability to be excited at any given temperature, compared to the electrons in QDs. As a consequence, the substrate ions deformability is smaller due to smaller electronic excitation probability, which in turn, explains why the lattice vibrational frequency for the QDs is smaller than the substrate~\cite{arul1}. This means that, the ions in the QDs are less rigid compared to their substrate. As anticipated, our results again agree qualitatively with the results calculated and measured experimentally in Refs.~\cite{zib}. 

Unlike the GaAs QDs discussed earlier, the reduced mass effect for InAs QDs also further supports the conclusion that $w_{\rm{ph}}^{\rm{SUB}}$ $>$ $w_{\rm{ph}}^{\rm{QD}}$. In other words, for InAs/GaAs system, we find that $M_r^{\rm{SUB}}$ (36.11) $<$ $M_r^{\rm{QD}}$ (45.34) $\rightarrow$ $w_{\rm{ph}}^{\rm{SUB}}$ $>$ $w_{\rm{ph}}^{\rm{QD}}$, which eventually satisfies the inequality, $t^{\rm{SUB}}$ $<$ $t^{\rm{QD}}$. One can surmise here that there is no competition between the ionization energy effect and the reduced mass contribution to the phonons. 

\subsection{Further analysis}
    
At this juncture, one may wonder two possibilities, namely, (1) if the size of the QDs are much smaller than the substrate, then the inequality, $\xi^{\rm{TOT}}_{\rm{SUB}}$ $>$ $\xi^{\rm{TOT}}_{\rm{QD}}$ as stated for InAs QDs may not be true for all QD sizes, and as a result of this, (2) there is a possibility to grow dots that are not electronically isolated, in other words, during growth, the dots at a certain small size may satisfy $\xi^{\rm{TOT}}_{\rm{SUB}}$ $=$ $\xi^{\rm{TOT}}_{\rm{QD}}$, or even $\xi^{\rm{TOT}}_{\rm{SUB}}$ $<$ $\xi^{\rm{TOT}}_{\rm{QD}}$ for very small InAs QDs. Hence, one can indeed in principle grow dots with zero confinement energy by choosing the appropriate QD size and elemental composition for a given substrate. However, the changes in ionization energy due to size is much less compared to the change due to elemental composition (between InAs QDs and GaAs matrix). In other words, the inequality, $\xi^{\rm{TOT}}_{\rm{SUB}}$ $>$ $\xi^{\rm{TOT}}_{\rm{QD}}$ will not be physically reversible for InAs QDs surrounded by GaAs matrix by only changing the QD size because of large contribution due to elemental composition (1.5 eV atom$^{-1}$). To see this effect clearly, let us write down the respective ionization energy approximation equations for the InAs/GaAs system, which can be obtained from (using Eq.~(8) given in Ref.~\cite{arul4})

\begin {eqnarray}
&&E_0 \pm \xi = E_{\rm{kinetic}} + V_{\rm{Coulomb}} + V_{\rm{body}}^{\rm{many}} \nn \\&& = E_0 \pm \beta \sum_i^z \frac{E_{Ii}}{z}. \label{eq:4b}
\end {eqnarray}

where 

\begin {eqnarray}
\beta = 1 + \frac{\langle V_{\rm{body}}^{\rm{many}}\rangle}{E_I}. \label{eq:5b}
\end {eqnarray}

Consequently, we can identify $\langle V_{\rm{body}}^{\rm{many}}\rangle$ as the atomic screened Coulomb potential by assuming that QDs are artificial atoms, which is given by~\cite{arul1} (we apply Eq.~(\ref{eq:2}) to two-electron hydrogen-like atomic systems)

\begin {eqnarray}
\hat{H} = \hat{H}_{\rm{o}} + \hat{V}_{\rm{sc}}, \label{eq:1a}
\end {eqnarray}   

and its solution is given by~\cite{arul1}

\begin {eqnarray}
\hat{\left\langle H\right\rangle} = 2Z^2E_1 + \hat{\left\langle V\right\rangle}_{\rm{sc}} = 2Z^2E_1 + \frac{40Z^6E_1}{\big[2Z+a_B\sigma]^5}. \label{eq:11a} 
\end {eqnarray}

Where, $Z$ is the atomic number, 

\begin {eqnarray}
\hat{V}_{\rm{sc}} = \frac{e}{4\pi\epsilon_0r}e^{-\mu re^{\frac{1}{2}\lambda(- \xi)}} = \frac{e}{4\pi\epsilon_0r}e^{-\sigma r}. \label{eq:11c}
\end {eqnarray}

Here, $\mu$ is the screening parameter's constant of proportionality and 

\begin {eqnarray}
E_1 = -\left[\frac{m}{2\hbar^2}\bigg(\frac{e^2}{4\pi\epsilon_0}\bigg)^2\right]. \label{eq:12a} 
\end {eqnarray}

Using Eqs.~(\ref{eq:1a}) and~(\ref{eq:11a}), we can surmise that indeed smaller $\xi_{\rm bulk}$ gives rise to smaller confinement. Now, the change in $\xi_{\rm bulk}$ due to elemental composition is $\delta_{\rm{bulk}}^{\xi} = \xi^{\rm{SUB}}_{\rm{bulk}} - \xi^{\rm{QD}}_{\rm{bulk}}$. Whereas, the change in the $\xi$ due to size is $\delta_{\rm{size}}^{\xi} = \xi^{\rm{SUB}}_{\rm{size}} - \xi^{\rm{QD}}_{\rm{size}}$. Since the change in the ionization energy due to size is smaller compared to elemental composition, we can assume that $|\delta_{\rm{bulk}}^{\xi}|$ $>$ $|\delta_{\rm{size}}^{\xi}|$, which was done earlier. Even if $|\delta_{\rm{bulk}}^{\xi}|$ $>$ $|\delta_{\rm{size}}^{\xi}|$ is not true, then it is possible to grow very small QDs that are electronically coupled to the substrate. Recall here that this electronic coupling means that the electronic excitation probabilities for both QDs and the substrate are identical. 

In summary, using the IET, it is possible for us to select suitable substrate and QD materials (elemental composition) in order to switch-off the contribution of substrate phonons. This will help the experimenters to grow QDs with specific application in mind, be it for quantum computing (with negligible phonon contribution from substrate) or photovoltaic cells (with large confinement energy and small $e$-$ph$ scattering). 

\begin{figure}[hbtp!]
\begin{center}
\scalebox{0.3}{\includegraphics{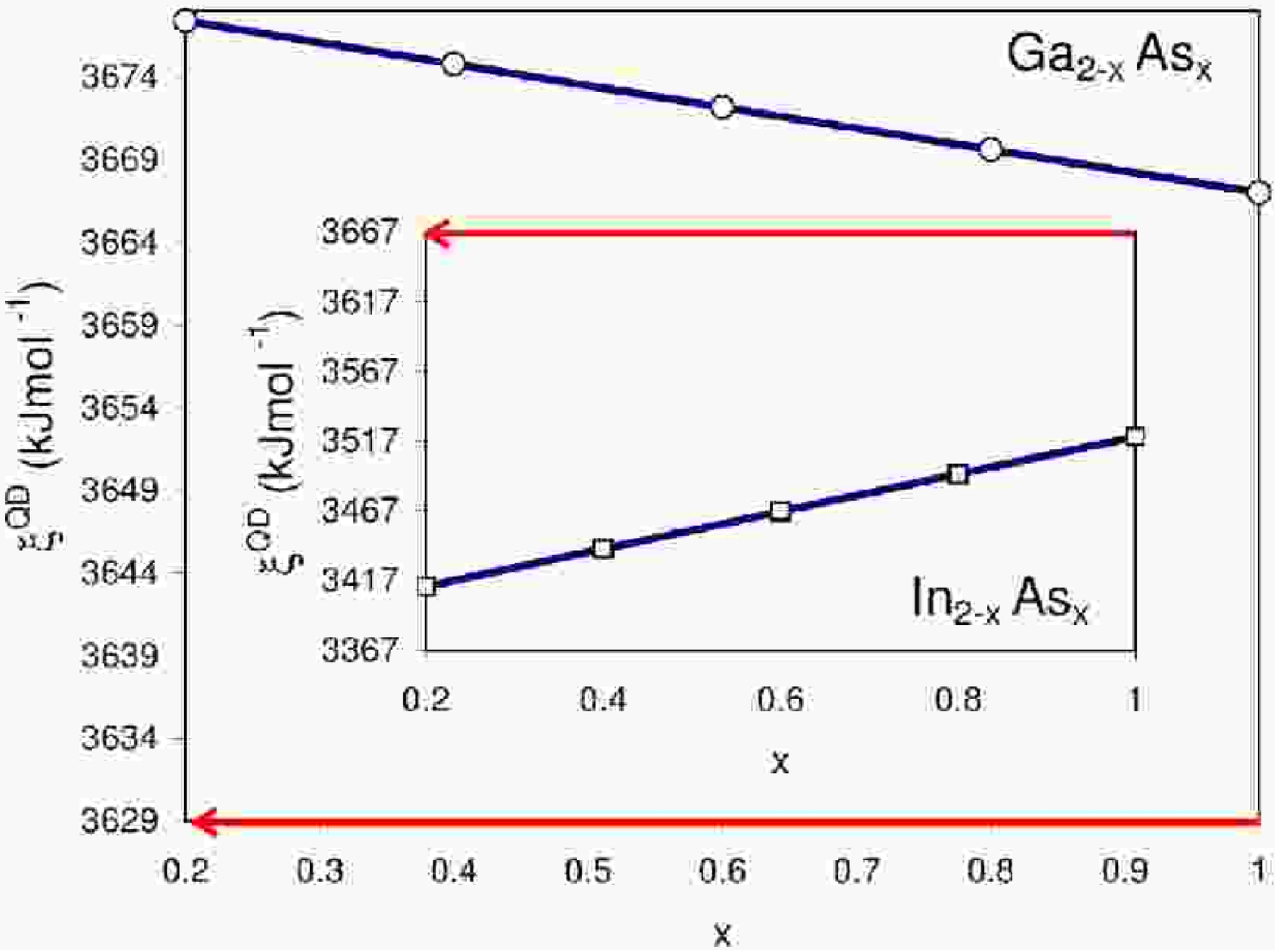}}
\caption{The evolution of the QD's ionization energy with respect to systematic substitutional doping for Ga$_{2-x}$As$_x$/Al$_{0.3}$Ga$_{0.7}$As and In$_{2-x}$As$_x$/GaAs (Inset). Substituting atomic Ga with As decreases the total ionization energy of the QDs (Ga$_{2-x}$As$_x$) but it is still larger than the total ionization energy of the substrate (Al$_{0.3}$Ga$_{0.7}$As), which is 3629 kJmol$^{-1}$ (follow the arrow). INSET: The total ionization energy for the In$_{2-x}$As$_x$ QDs increases with $x$ but it will still be smaller than the total ionization energy of the substrate (3667 kJmol$^{-1}$, follow the arrow), regardless of $x$. Therefore, both QDs are electronically isolated from their respective substrates for all doping, $x$. See text for details.}
\label{fig:3}
\end{center}
\end{figure}

For example, Fig.~\ref{fig:3} indicates what to expect if we were to change the doping in Ga$_{2-x}$As$_x$ and In$_{2-x}$As$_x$ QDs systematically. Regardless of doping $x$, we will find that the total ionization energy of the Ga$_{2-x}$As$_x$ QDs will always be larger than their substrate's value. The reason is that the total ionization energies for both $x = 2$ and $x = 0$ are still larger than their substrate's averaged value, 3629 kJmol$^{-1}$. On the other hand, the total ionization energies for In$_{2-x}$As$_x$ QDs for both $x = 2$ and $x = 0$ are smaller than their substrate's value, which is 3667 kJmol$^{-1}$ (see the inset of Fig.~\ref{fig:3}). Consequently, these QDs will always be electronically isolated regardless of the doping $x$. In other words, Ga$_{2-x}$As$_x$ QDs will remain adiabatically independent from their substrate for all doping $x$ (pure dephasing). Whereas, for In$_{2-x}$As$_x$ QDs, the substrate phonons will keep influencing the electrons in the QDs for all doping $x$ (inelastic dephasing). It is important to note here that even if we were to increase the size of these QDs to form thin films, these thin films will still be electronically isolated from their substrate due to their different elemental composition. That is why thin films of different elemental composition from their substrate will have different electronic properties compared to their substrate. This will remain true for as long as the elemental composition between the substrate and the thin films are different.

\section{Dephasing mechanisms and ionization energy}

Here, we discuss the various types of dephasing rates (relaxation lifetimes, $\tau$) due to defects (def), impurities (imp: magnetic and non-magnetic), spin, $e$-$e$ and $e$-$ph$ interactions, and how they are related to the IET. These effects are first analyzed in the absence of any external perturbations such as laser (to create excitons) and applied magnetic fields. Subsequently, further evaluations are carried out based on the changes in the spin transition probability by identifying the fact that the inelastic dephasing is favorable with increasing transition probability between the spin-down and spin-up states. Whereas, pure dephasing is preferred if the transition probability approaches zero. Hence, our strategy here is to relate the changes in the spin transition probability with the IET, and subsequently to the inelastic or pure dephasing mechanism. Apparently, we cannot use this strategy to calculate the dephasing rates. However, the changes in the transition probability can be related to the dephasing rates. This means that, if the transition probability approaches one, then the dominant dephasing rate is determined by the inelastic dephasing. 

In contrast, in the event where the transition probability approaches zero, then the pure dephasing will act as the dominant mechanism and determines the dephasing rate. Importantly, the ionization energy approximation requires that all the electrons are correlated and they are not free-electrons, be it in the conduction or valence band. Thus, this approach is not suitable for free-electron metallic substrates. Recall that our strategy here is to develop a theory to understand why the experimental data in Ref.~\cite{sang} is different from Ref.~\cite{zib}. 

\subsection{Dephasing rates}

First, the total dephasing rate (relaxation lifetime) for QDs can be written as~\cite{bala}

\begin {eqnarray}
&&\frac{1}{\tau^{\rm{QD}}} = \frac{1}{\tau_{\rm{e-e}}} + \frac{1}{\tau_{\rm{e-ph(SUB)}}} + \frac{1}{\tau_{\rm{e-ph(QD)}}} \nonumber \\&& + \frac{1}{\tau_{\rm{imp}}} + \frac{1}{\tau_{\rm{def}}} + \frac{1}{\tau_{\rm{spin}}}, \label{eq:A1} 
\end {eqnarray}  

where, $\tau^{\rm{QD}}$ is the total dephasing rate in a QD, while the other respective dephasing rates are due to $e$(QD)-$e$(QD), $e$(QD)-$ph$(SUB) and $e$(QD)-$ph$(QD) interactions, impurities, defects and spin. For non-magnetic and/or magnetic system above the Kondo temperature and in the absence of applied magnetic field, we can ignore $\tau_{\rm{spin}}$, while $\tau_{\rm{imp}}$ and $\tau_{\rm{def}}$ can be re-labeled as $\tau_{\rm{D}}$ that includes all types of defects (vacancy, interstitial, substitution at different sites), including different types of elements. Thus, Eq.~(\ref{eq:A1}) can now be rewritten as 

\begin {eqnarray}
\frac{1}{\tau^{\rm{QD}}} = \frac{1}{\tau_{\rm{e-e}}} + \frac{1}{\tau_{\rm{e-ph(SUB)}}} + \frac{1}{\tau_{\rm{e-ph(QD)}}} + \frac{1}{\tau_{\rm{D}}}. \label{eq:A2} 
\end {eqnarray}         

The next step is to invoke Eqs.~(\ref{eq:25}) and~(\ref{eq:26}), which imply that the electrons in the QD (e(QD)) are influenced by the phonons from the QD (ph(QD)) and substrate (ph(SUB)). Therefore, $\tau_{\rm{e-e}}$, $\tau_{\rm{e-ph(SUB)}}$ and $\tau_{\rm{e-ph(QD)}}$ can be simply written as 

\begin {eqnarray}
\frac{1}{\tau^{\rm{QD}}} = \frac{1}{\tau^{\rm{e(QD)}}_{\rm{ph(SUB)}}} + \frac{1}{\tau^{\rm{e(QD)}}_{\rm{ph(QD)}}}. \label{eq:A3} 
\end {eqnarray} 
    
The reasoning used to arrive at Eq.~(\ref{eq:A3}) implies that the effects of defects, impurities, and $e$-$e$ interaction in the QDs and substrate have been captured via e(QD), ph(SUB) and ph(QD). In other words, we can write Eq.~(\ref{eq:A1}) in the form of Eq.~(\ref{eq:A3}) because any systematic changes to these effects also affect the valence states (of each element) and/or the ionization energy of the system systematically. All our discussion in the previous sections are based on Eq.~(\ref{eq:A3}). For example, longer $\tau^{\rm QD}$ is only possible if $t^{\rm SUB,QD} > \tau^{\rm QD}$ ($\xi^{\rm QD} > \hbar w_{\rm ph}^{\rm SUB,QD}$): adiabatically decoupled and pure dephasing as the dominant mechanism, or $t^{\rm SUB,QD} < \tau^{\rm QD}$ ($\xi^{\rm QD} < \hbar w_{\rm ph}^{\rm SUB,QD}$): adiabatically coupled and inelastic dephasing as the dominant mechanism.

\subsection{Spin transition probability and the relaxation lifetime}

\textbf{Claim 3}: Even though we did not calculate the total dephasing rate in QDs explicitly under different perturbations (laser, electric and magnetic fields) using Eq.~(\ref{eq:A3}), but we have invoked that the transition probability of a confined electron in a QD from its ground to the excited state is proportional to the inverse external timescales, $1/t^{\rm{SUB}}$ and $1/t^{\rm{QD}}$. For example, $t^{\rm{SUB}} > \tau^{\rm{QD}}$ and $t^{\rm{QD}} > \tau^{\rm{QD}}$ give rise to pure dephasing, while $t^{\rm{SUB}} < \tau^{\rm{QD}}$ and $t^{\rm{QD}} < \tau^{\rm{QD}}$ lead to inelastic dephasing as discussed earlier. 

\textbf{Proof 3}: Here we will prove why such proportionality (transition probability $\propto$ $1/t^{\rm{SUB,QD}}$) is valid by means of the well known hamiltonian of an electron that starts out as a spin-up electron in the presence of a rotating ($\omega$) magnetic field ($B_0$) at an angle, $\alpha$. The hamiltonian and its transition probability to spin down is given by~\cite{gri}     

\begin {eqnarray}
&&H(t) = \frac{\hbar \omega_1}{2} [A], \label{eq:A4} \\&& A = \bigg[\sin\alpha\cos(\omega t)\sigma_x + \sin\alpha\sin(\omega t)\sigma_y + \cos\alpha\sigma_z\bigg], \nn\\&& \left|\left\langle \chi(t)|\chi_-(t)\right\rangle\right|^2 = \bigg[\frac{\omega}{\lambda}\sin\alpha\sin\bigg(\frac{\lambda t}{2}\bigg)\bigg]^2, \label{eq:A6}
\end {eqnarray}
      
where, $\sigma_{x}$, $\sigma_{y}$ and $\sigma_{z}$ are the Pauli spin matrices, $\lambda = \sqrt{\omega^2 + \omega_1^2 - 2\omega\omega_1\cos\alpha}$, and $\chi_{\pm}(t)$ denote the normalized eigenspinors. Here,

\begin {eqnarray} 
&&\chi(t) = \bigg[\cos(\lambda t/2) - i\frac{(\omega_1 - \omega\cos\alpha)}{\lambda}\sin(\lambda t/2)\bigg] \nn \\&& \times e^{-\frac{i\omega t}{2}}\chi_{+}(t)  + i\bigg[\frac{\omega}{\lambda}\sin\alpha\sin(\lambda t/2)\bigg]e^{\frac{i\omega t}{2}}\chi_{-}(t), \label{eq:new} \\&&
\chi_+(t) = (\cos(\alpha/2), e^{i\omega t}\sin(\alpha/2)),\nn \\&&
\chi_-(t) = (e^{-i\omega t}\sin(\alpha/2), -\cos(\alpha/2)).\nn 
\end {eqnarray}

The angular velocity, $\omega = 1/T_{\rm{ext}}$ refers to the characteristic time for the change in the Hamiltonian (external) given in Eq.~(\ref{eq:A4}), while $\omega_1 = eB_0/m = 1/T_{\rm{int}}$ refers to the characteristic time for the changes in the wave function (internal). 

\begin{figure}[hbtp!]
\begin{center}
\scalebox{0.3}{\includegraphics{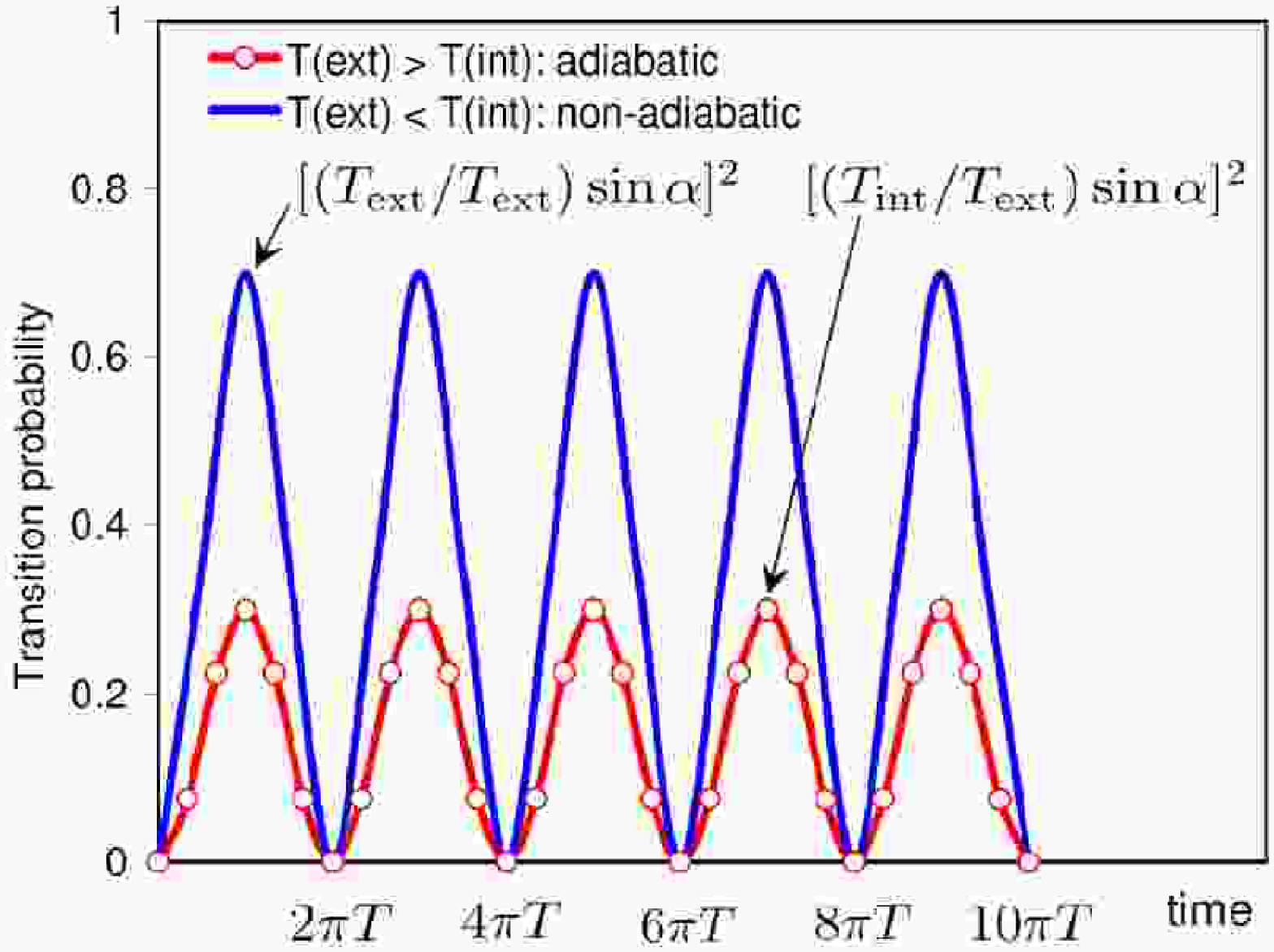}}
\caption{Spin transition probability from spin-up to spin-down based on Eq.~(\ref{eq:A6}). The label, $T_{\rm{ext}} > T_{\rm{int}}$ indicates the system approximately satisfies the adiabatic condition, while $T_{\rm{ext}} < T_{\rm{int}}$ implies the spins are non-adiabatically coupled to the external magnetic field ($B_0$). Thus, $1/T_{\rm ext}$ determines the spin transition probability (spin-up to spin-down). These plots have peaks at $[(T_{\rm{ext}}/T_{\rm{ext}})\sin\alpha]^2$ and $[(T_{\rm{int}}/T_{\rm{ext}})\sin\alpha]^2$ for $T_{\rm{ext}} < T_{\rm{int}}$ and $T_{\rm{ext}} > T_{\rm{int}}$, respectively. The timescale, $T$ labeled with $2\pi n T$, $n = 0, 1, ...$, in the time$-$axis corresponds to $T_{\rm{ext}}$ and $T_{\rm{int}}$ for $T_{\rm{ext}} < T_{\rm{int}}$ and $T_{\rm{ext}} > T_{\rm{int}}$, respectively.}
\label{fig:4}
\end{center}
\end{figure}

The transition probability given in Eq.~(\ref{eq:A6}) is plotted in Fig.~\ref{fig:4} that captures the anticipated proportionality between the transition probability and $1/T_{\rm{ext}}$. Recall here that the spin's timescale is $T_{\rm{int}}$. The plot labeled with $T_{\rm{ext}} > T_{\rm{int}}$ in Fig.~\ref{fig:4} implies adiabatic condition, in which, the spin transition probability is approximately independent of the external magnetic field, $B_0$ ($T_{\rm{ext}}$). Whereas, for $T_{\rm{ext}} < T_{\rm{int}}$, there is a non-adiabatic coupling between the external $B_0$ and the spin transition probability in the system. This in turn implies that $1/T_{\rm{ext}}$ $\propto$ $|\langle \chi(t)|\chi_-(t)\rangle |^2$, as claimed earlier $\blacksquare$ 

In other words, we have a clear indication that smaller $T_{\rm{ext}}$ (compared to $T_{\rm{int}}$) leads to larger $1/T_{\rm{ext}}$, which gives rise to larger spin transition probability (from spin-up to spin-down that corresponds to inelastic dephasing). On the other hand, large $T_{\rm{ext}}$ will lead us to smaller spin transition probability (this will enhance the pure dephasing mechanism). Using Eq.~(\ref{eq:A6}), we can calculate the peak value for the plot, $T_{\rm{ext}} < T_{\rm{int}}$ (non-adiabatic), which is given by, $[(T_{\rm{ext}}/T_{\rm{ext}})\sin\alpha]^2$. The peak value for the $T_{\rm{ext}} > T_{\rm{int}}$ (adiabatic) plot is $[(T_{\rm{int}}/T_{\rm{ext}})\sin\alpha]^2$. Therefore, the values for these peaks satisfy $[(T_{\rm{ext}}/T_{\rm{ext}})\sin\alpha]^2 > [(T_{\rm{int}}/T_{\rm{ext}})\sin\alpha]^2$ (see Fig.~\ref{fig:4}). 

Note here that the frequency for both plots are kept constant for graphical convenience by using the same magnitude for $T_{\rm{ext}}$ and $T_{\rm{int}}$ where the magnitudes of $T_{\rm{ext}}$ and $T_{\rm{int}}$ are interchanged when the inequality in $T_{\rm{ext}} < T_{\rm{int}}$ is changed from $<$ to $>$. In summary, we have shown that the transition probability is proportional to the external timescale $1/T_{\rm{ext}}$, which has been correctly invoked to discuss the experimental data in the earlier sections. For example, $1/T_{\rm{ext}}$ (related to rotating ($\omega$) magnetic field) can be identified as the $1/t^{\rm{SUB}}$ or $1/t^{\rm{QD}}$, which are related to the $w_{\rm{ph}}^{\rm{SUB}}$ and $w_{\rm{ph}}^{\rm{QD}}$, respectively in Eq.~(\ref{eq:24}). On the other hand, $1/T_{\rm{int}}$ ($\omega_1$) is related to the $1/\tau^{\rm{QD}}$ ($w_{\rm e}^{\rm QD}$).   

\section{Conclusions}

In this work, several electronic and phononic issues in quantum dots are discussed theoretically (as listed below), which also agree with recent experimental results.   
 
$\bullet$ The Green's and Gauss theorems have been invoked to derive the required electronic isolation of the quantum dots from the substrate in order to justify the electronic confinement. 

$\bullet$ Three conditions, (i), (ii) and (iii) were introduced to explain the physical mechanism required to isolate the quantum dots electronically. These conditions were used to discuss the phononic isolation that could occur in quantum dots, for different doping elements in QDs and substrates.    

$\bullet$ The ionization energy and its approximation have been employed to quantitatively explain how one could isolate the quantum dots from their substrate and/or matrix by varying their elemental compositions. 

$\bullet$ When the quantum dots are adiabatically isolated from the substrate, the pure dephasing process turns to play the leading role in determining its relaxation lifetime. On the other hand, if the phonons from the substrate are non-adiabatically coupled to the quantum dots, then the inelastic dephasing becomes the dominant mechanism. 

$\bullet$ All the theoretical results presented here, which are based on the ionization energy approximation and quantum adiabatic theorem agree well with the recently reported experimental and theoretical results.

Finally, the ionization energy method presented here can be used to evaluate the types of quantum dots that one can grow, whether they are isolated with respect to substrate electrons and phonons.  

\section*{Acknowledgments}

A.D.A. would like to thank the School of Physics, University of Sydney for the USIRS award and Micheal Delanty for pointing out Refs.~\cite{kaya,le} on QD shape effect. A.D.A. also thanks Kithriammah Soosay for the support. K.O. acknowledges the partial support from the Australian Research Council (ARC) and the CSIRO.

\end{document}